\documentstyle [11pt]{article}
\makeatletter

\newcommand{\singlespacing}{\let\CS=\@currsize\renewcommand{\baselinestretch}{1}\tiny\CS}
\newcommand{\oneandahalfspacing}{\let\CS=\@currsize\renewcommand{\baselinestretch}{1.25}\tiny\CS}
\newcommand{\doublespacing}{\let\CS=\@currsize\renewcommand{\baselinestretch}{1.35}\tiny\CS}

\def\@citex[#1]#2{\if@filesw\immediate\write\@auxout{\string\citation{#2}}\fi
  \def\@citea{}\@cite{\@for\@citeb:=#2\do
    {\@citea\def\@citea{,\linebreak[0]\hskip0pt plus .2em}%
      \@ifundefined{b@\@citeb}%
      {{\bf ?}\@warning{Citation `\@citeb' on page \thepage\space undefined}}%
      \hbox{\csname b@\@citeb\endcsname}}}{#1}}

\newtheorem{rule-def}[theorem]{Rule}

\begin{document}
%\setcounter{chapter}{1}
% defining short form------
\newcommand{\la}{\lambda}
\newcommand{\si}{\sigma}
\newcommand{\ol}{1-\lambda}
\newcommand{\be}{\begin{equation}}
\newcommand{\ee}{\end{equation}}
\newcommand{\bea}{\begin{eqnarray}}
\newcommand{\eea}{\end{eqnarray}}
\newcommand{\nn}{\nonumber}
\newcommand{\lb}{\label}

\begin{center}
{\large \bf RELATIVISTIC ELECTROMAGNETIC MASS MODELS: CHARGED DUST
DISTRIBUTION IN HIGHER DIMENSIONS}
\end{center}

\begin{center}
SAIBAL RAY$^1$, SUMANA BHADRA$^2$ and G. MOHANTI$^3$\\
 $^1${\it Department of Physics, Barasat Government College, Barasat 743 201, North
 24 Parganas, West Bengal, India \& Inter-University Centre for Astronomy and
  Astrophysics, PO Box 4, Pune 411 007, India}\\
$^2${\it Balichak Girls' High School, Balichak 721 124, Midnapur,
West Bengal, India}\\ $^3${\it School of Mathematical Sciences,
Sambalpur University, Burla 768 019, Orissa, India}
\end{center}

{\bf Abstract.}\\
 Electromagnetic mass models are proved to exist
in higher dimensional theory of general relativity corresponding
to charged dust distribution. Along with the general proof a
specific example is also sited as a supporting candidate.\\
 PACS
number(s): 04.20.-q, 04.40.Nr, 04.20.Jb\\

\begin{center}
{\bf 1. Introduction}
\end{center}
 In many theories higher dimensions play an
important role, specially in super string theory (Schwarz, 1985;
Weinberg, 1986) which demands more than usual four dimensional
space-time. This is also true in studying the models regarding
unification of gravitational force with other fundamental forces
in nature. In the case of a simple solution to the vacuum field
equations of general relativity in $4+1$ space-time dimensions
Chodos and Detweiler (1980) have shown that it leads to a
cosmology which at the present epoch has $3+1$ observable
dimensions in which the Einstein-Maxwell equations are obeyed.
Lorenz-Petzold (1984) has studied a class of higher dimensional
Bianchi-Kantowski-Sachs space-times of the Kaluza-Klein type
whereas Ibanez and Veraguer (1986) have examined radiative
isotropic cosmologies with extra dimensions related to FRW models.

In this connection it is interesting to note that electromagnetic
mass models where all the physical parameters, including the
gravitational mass, are arising from the electromagnetic field
alone have been extensively studied (Tiwari et al., 1984;
Gautreau, 1985; Gr{\o}n, 1986; Ponce de Leon, 1987; Tiwari and
Ray, 1991; Ray and Bhadra, 2004) in the space-time of four
dimensional general relativity. Thus it is believed that study of
electromagnetic mass models in higher dimensional theory will be
physically more interesting.

Under this motivation we have considered here a static spherically
symmetric charged dust distribution corresponding to
higher dimensional theory of general relativity. It is proved, as a
particular case, from the
coupled Einstein-Maxwell field equations that a bounded and regular
interior static spherically
symmetric charged dust solution, if exists, can only be of purely
electromagnetic origin. An example,
which is already available, is examined in this context and is shown
that the solution set satisfies
the condition of being electromagnetic origin.

\begin{center}
{\bf 2. The Einstein-Maxwell Field Equations}
\end{center}
The Einstein-Maxwell field equations for the case of charged dust distribution are given by
\begin{eqnarray}
{G^{i}}_{j} = {R^{i}}_{j} - \frac{1}{2}{{g^{i}}_{j}} R = -8\pi [{{T^{i}}_{j}}^{(m)}+{{T^{i}}_{j}}^{(em)}],
\end{eqnarray}
\begin{eqnarray}
{[{(-g)}^{1/2}F^{ij}], }_{j}= 4\pi J^{i}{(-g)}^{1/2},
\end{eqnarray}
\begin{eqnarray}
F_{[ij, k]}= 0
\end{eqnarray}
\noindent
where ${F^{ij}}$ is the electromagnetic field tensor and ${J^{i}}$ the current four
vector which is equivalent to
\begin{eqnarray}
{J^{i}}= \sigma {u^{i}}
\end{eqnarray}
$\sigma $ being the charge density and $u^{i}$ is the four velocity of the matter
satisfying the relation
\begin{eqnarray}
u_{i}u^{i}= 1.
\end{eqnarray}
The matter and electromagnetic energy momentum tensors are, respectively, given by
\begin{eqnarray}
{{T^{i}}_{j}}^{(m)}= \rho u^{i}u_{j},
\end{eqnarray}
\begin{eqnarray}
{{T^{i}}_{j}}^{(em)}= \frac{1}{4\pi}[- F_{jk}F^{ik} +\frac{1}{4}{g^{i}}_{j}F_{kl}F^{kl}],
\end{eqnarray}
where $\rho $ is the proper energy density.\\
\noindent
  Now we consider the $(n+2)$ dimensional spherically symmetric metric
\begin{eqnarray}
ds^{2}= e^{\nu(r)}dt^{2}- e^{\lambda(r)}dr^{2}- r^{2}{X_{n}}^{2}
\end{eqnarray}
where
\begin{eqnarray}
{X_{n}}^{2}=
d{\theta_{1}}^{2}+sin^{2}{\theta_{1}}d{\theta_{2}}^{2}+\nonumber
sin^{2}{\theta_{1}}sin^{2}{\theta_{2}}d{\theta_{3}}^{2}+......\nonumber
+\left[\prod^{n-1}_{i=1}sin^{2}{\theta_{i}}\right]d{\theta_{n}}^{2}.
\end{eqnarray}
The convention adopted here for coordinates are
\begin{eqnarray}
x^{1}= r,\quad x^{2}=\theta_{1},\quad x^{3}=\theta_{2},.....x^{n+1}=\theta_{n},\quad x^{n+2}= t
\end{eqnarray}
and also
\begin{eqnarray}
g_{11}=-e^\lambda,\quad g_{22}= -r^{2},\quad g_{33}=-r^{2}sin^{2}{\theta_{1}},\nonumber\\
g_{44}=-r^{2}sin^{2}{\theta_{1}}sin^{2}{\theta_{2}},..... g_{(n+1)(n+1)}=-r^{2}\left[\prod^{n-1}_{i=1}{sin^{2}{\theta_{i}}}\right],\nonumber\\
g_{(n+2)(n+2)}= e^\nu.
\end{eqnarray}
\noindent
As we have considered here a static fluid, so
\begin{eqnarray}
u^{i}= [0,0,0,.....(n+1)times,\quad e^{-{\nu}/{2}}],
\end{eqnarray}
\begin{eqnarray}
J^{1}= J^{2}= J^{3}=...J^{n+1}= 0,\quad J^{n+2}\neq 0
\end{eqnarray}
so that the only non-vanishing components of $F_{ij}$ of equations
${2}$ and ${3}$ are $F_{{1}(n+2)}$ and $F_{(n+2){1}}$.\\ \noindent
In view of the above, the Einstein-Maxwell field equations for
static spherically symmetric charged dust corresponding to the
metric (8) are
\begin{eqnarray}
e^{-\lambda}[n{\nu^{\prime}/{2r}}+n(n-1)/2r^{2}]-n(n-1)/2r^{2}= -E^{2},
\end{eqnarray}
\begin{eqnarray}
e^{-\lambda}[\nu^{{\prime}{\prime}}/2+{\nu^{\prime}}^{2}/4-
{\nu^{\prime}\lambda^{\prime}}/4+(n-1)(\nu^{\prime}-\lambda^{\prime})/2r+\nonumber \\
(n-1)(n-2)/2r^{2}] -(n-1)(n-2)/2r^{2}= E^{2},
\end{eqnarray}
\begin{eqnarray}
e^{-\lambda}[n{\lambda^{\prime}}/2r-n(n-1)/2r^{2}]+n(n-1)/2r^{2}= 8\pi\rho+E^{2},
\end{eqnarray}
\begin{eqnarray}
[r^{n}E]^{\prime}= 4\pi r^{n} \sigma e^{\lambda/2}
\end{eqnarray}
where $E$, the electric field strength, is defined as
$ E= -e^{-(\nu+\lambda)/2}\phi^\prime$ ,
the electrostatic potential  $\phi$  being related to the
electromagnetic field tensor as $F_{(n+2)1}= -F_{1(n+2)}=\phi^{\prime}$.

\begin{center}
{\bf 3. Higher Dimensional Electromagnetic Mass Models}
\end{center}
From the field equations (13) and (15), we have
\begin{eqnarray}
e^{-\lambda}(\nu^\prime+\lambda^\prime)= 16\pi r\rho/n.
\end{eqnarray}
Now, we make use of the conservation equations ${{T^{i}}_{j}};i= 0$ which yield
\begin{eqnarray}
\rho{\nu^\prime}= [q^{2}]^\prime/4\pi r^{4} + (n-2)E^2/2\pi r
\end{eqnarray}
where the charge, $q$, is related with the electric field
strength, $E$, through the integral form of the Maxwell's equation
(16), which can be written as
\begin{eqnarray}
q = Er^{n} = 4\pi\int_{0}^{r}\sigma e^{\lambda/2}r^{n}dr.
\end{eqnarray}
Again,  equation (15) can be expressed in the following form as
\begin{eqnarray}
e^{-\lambda}= 1- 4M/nr^{n-1}
\end{eqnarray}
where the active gravitational mass, $M$, is given by
\begin{eqnarray}
M = 4\pi\int_{0}^{r}[\rho+E^{2}/8\pi]r^{n}dr.
\end{eqnarray}
Hence, following the technique of Tiwari and Ray (1991) we see
that for vanishing charge density, $\sigma$, via  equation (19)
one gets from  equation (18) the unique relation
\begin{eqnarray}
\rho\nu^\prime= 0.
\end{eqnarray}
Thus, we have the following two cases:\\ \noindent {\it Case I:
$\rho \ne0$, \quad $\nu^\prime= 0$}\\ \noindent For this case,
from equations (13) and (14), we have $\lambda$ to be a constant.
This in turn makes $\rho$ equal to zero and hence by virtue of
equation (22) space-time becomes flat.\\ \noindent {\it Case II:
$\rho=0$,\quad $\nu^\prime\ne0$}\\ \noindent  In this case, from
equations (20) and (21), one can see that $\lambda$ becomes zero.
Then from equation (17), we have the metric potential as a
constant and again the space-time becoming flat.

We are not considering here the third case, viz. $\rho=
\nu^\prime= 0$, which is quite a trivial one. However, from the
above cases (i) and (ii) it is evident that, at least at a
particular case, all the charged dust models are of
electromagnetic origin, viz., all the physical parameters
originating purely from electromagnetic field. This type of models
are known as electromagnetic mass models in the literature
(Lorentz, 1904; Feynman, 1964).

{\bf  An example}:\\ The solution set obtained by Khadekar et al.
(2001) for the static spherically symmetric charged dust is as
follows:
\begin{eqnarray}
e^\nu = Ar^{2N},
\end{eqnarray}
\begin{eqnarray}
e^{-\lambda} =\left[\frac{(n-1)}{N+(n-1)}\right]^{2},
\end{eqnarray}
\begin{eqnarray}
\rho =\frac{Nn(n-1)^{2}}{8 \pi r^{2}[N+(n-1)]^{2}},
\end{eqnarray}
\begin{eqnarray}
\sigma = \frac{N(n-1)^{2}[n(n-1)]^{1/2}}{4 \pi 2^{1/2}r^2[N+(n-1)]^2}.
\end{eqnarray}
where $A$ and $N$ both are constants with the restriction that $N\geq0$.\\
The total charge and mass of the sphere in terms of its radius, $a$,
are respectively given by
\begin{eqnarray}
q = \frac{N[n(n-1)]^{1/2}a^{n-1}}{2^{1/2}[N+(n-1)]},
\end{eqnarray}
\begin{eqnarray}
m =\frac{Na^{n-1}}{N+(n-1)}.
\end{eqnarray}
The charge and mass densities in the present case take the relationship
\begin{eqnarray}
\sigma = [2(1-1/n)]^{1/2}\rho.
\end{eqnarray}
Therefore, the charge and mass densities are proportional to each
other with the constant of proportionality $[2(1-1/n)]^{1/2}$
which takes the value unity for $n=2$ i.e. in the four dimensional
case and the relation (29) reduces to the usual form
\begin{eqnarray}
\sigma = \pm\rho
\end{eqnarray}
which is known as the De-Raychaudhuri (1968) condition for
equilibrium of a charged fluid.\\ Thus, from equations (23)-(29),
it is evident that all the physical quantities including the
effective gravitational mass vanish and also the spherically
symmetric space-time becomes flat when the charge density vanishes
implying $N=0$. The solution here, therefore, satisfies the
criteria of being of purely electromagnetic origin.

\begin{center}
{\bf 4. Discussions}
\end{center}
The present paper is, in general, higher dimensional analogue of
the work of Tiwari and Ray (1991) whereas the example given here
(Khadekar et al., 2001) is the higher dimensional analogue of the
paper of Pant and Sah (1979). Thus, we have presented here a model
which corresponds to spherically symmetric gravitational sources
of purely electromagnetic origin in the space-time of higher
dimensional theory of general relativity. It is already proved in
the four dimensional presentation of the present paper (Tiwari and
Ray, 1991) that a bounded continuous static spherically symmetric
charged dust solution, if exists, can only be of electromagnetic
origin. Hence, this is also true in the higher dimension of
general theory of relativity.

In this regard we would like to discuss briefly the role of higher
dimensions in different context. It have been shown by Ibanez and
Veraguer (1986) that for the open models related to FRW
cosmologies the extra dimensions contract as a result of
cosmological evolution whereas for flat and closed models they
contract only when there is one extra dimension. Fukui (1987)
recovers Chodos-Detweiler (1980) type solutions, as mentioned in
the introduction, where the Universe expands as $t^{1/2}$ by the
percolation of radiation into 4D space-time from the fifth
dimension, mass, although the 5D space-time-mass Universe itself
is in vacuum as a whole. It is interesting to note that
considering mass as fifth dimension a lot of other works also have
been done by several researchers (Wesson 1983; Banerjee, Bhui and
Chatterjee 1990; Chatterjee and Bhui 1990) which contain
Einstein's theory embedded within it. In one of such
investigations it is argued that a huge amount of entropy can be
produced following shrinkage of extra-dimension which may account
for the very large value of entropy per baryon observed in 4D
world (Chatterjee and Bhui 1990). Kaluza-Klein type higher
dimenssional inflationary scenario have been discussed by Ishihara
(1984) and Gegenberg and Das (1985) where it is shown that the
contraction of the internal space causes the inflation of the
usual space.\\

\begin{center}
{\bf 4. Acknowledgments}
\end{center}
One of the authors (SR) would like to express his gratitude to the
authority of Inter-University Centre for Astronomy and
Astrophysics, PO Box 4, Pune 411 007, India for providing him
Associateship Programme under which a part of this work was
carried out. Support under UGC grant (No. F-PSN-002/04-05/ERO) is
gratefully acknowledged.

\pagebreak

\begin{center}
{\bf References}
\end{center}
\noindent Banerjee, A., Bhui, B. K. and Chatterjee, S.: 1990, {\it
Astron. Astrophys.} {\bf 232}, 305.\\ \noindent Chatterjee, S. and
Bhui, B. K.: 1990, {\it Astrophys. Space Sci.} {\bf 167}, 61.\\
\noindent Chodos, A. and Detweiler, S.: 1980, {\it Phys. Rev. D}
{\bf 21}, 2167.\\ \noindent De, U. K. and Raychaudhuri, A. K.:
1968, {\it Proc. Roy. Soc. London A} {\bf 303}, 97.\\ \noindent
Feynman, R. P., Leighton, R. R. and Sands, M.: 1964, {\it The
    Feynman Lectures in Physics} (Addison-Wesley, Palo Alto, Vol. II,
    Chap. 28).\\
\noindent Fukui, T.: 1987, {\it Gen. Rel. Grav.} {\bf 19}, 43.\\
\noindent Gautreau, R.: 1985, {\it Phys. Rev. D} {\bf 31}, 1860.\\
\noindent Gegenberg, J. D. and Das, A.: 1985, {\it Phys. Lett. A}
{\bf 112}, 427.\\ \noindent Gr{\o}n, {\O}.: 1986, {\it Gen. Rel.
Grav.} {\bf 18}, 591. \\ \noindent Ibanez, J. and Veraguer, E.:
1986, {\it Phys. Rev D} {\bf 34}, 1202.\\ \noindent Ishihara, H.:
1984, {\it Prog. Theor. Phys.} {\bf 72}, 376.\\ \noindent
Khadekar, G. S., Butey, B. P. and Shobhane, P. D.: 2001, {\it Jr.
Ind. Math. Soc.}, {\bf 68}, 33.\\ \noindent Lorentz, H. A.: 1904,
{\it Proc. Acad. Sci., Amsterdam 6} (Reprinted
  in Einstein et al., The Principle of Relativity, Dover, INC, 1952, p.
  24).\\
\noindent
  Lorenz-Petzold, D.: 1984, {\it Phys. Lett. B} {\bf 149}, 79.\\
   \noindent Pant, D. N. and Sah, A.:
1979, {\it J. Maths. Phys.} {\bf 20}, 2537. \\ \noindent Ponce de
Leon, J.: 1987, {\it J. Maths. Phys.} {\bf 28}, 410. \\ \noindent
Ray, S. and  Bhadra, S.: 2004, {\it Phys. Lett. A} {\bf 322},
150.\\
 \noindent
 Schwarz, J.H.: 1985,
{\it Superstings} (World Scientific, Singapore).\\
 \noindent
 Tiwari, R. N., Rao, J. R. and
Kanakamedala, R. R.: 1984, {\it Phys. Rev.D} {\bf 30}, 489.\\
\noindent
Tiwari, R. N. and Ray, S.: 1991, {\it Astrophys. Space
Sc.} {\bf 180}, 143.\\
 \noindent
Weinberg, S.: 1986, {\it Strings and Superstrings} (World
   Scientific, Singapore).\\
\noindent Wesson P. S.: 1983, {\it Astron. Astrophys.} {\bf 119},
145.\\
\end{document}